# A Universe of Sound: processing NASA data into sonifications to explore participant response

Kimberly Kowal Arcand[1†], Jessica Sarah Schonhut-Stasik[2,3*†], Sarah G. Kane[4,5†], Gwynn Sturdevant[6†], Matt Russo[7], Megan Watzke[1], Brian Hsu[8] and Lisa F. Smith[9]

[1]Department of High Energy Astrophysics, Smithsonian Astrophysical Observatory, Cambridge, MA, United States, [2]Department of Physics and Astronomy, Vanderbilt University, Nashville, TN, United States, [3]Neurodiversity Inspired Science and Engineering Fellow, Frist Center for Autism and Innovation, Vanderbilt University, Nashville, TN, United States, [4]Institute of Astronomy, University of Cambridge, Cambridge, United Kingdom, [5]Department of Physics and Astronomy, University of Pennsylvania, Philadelphia, PA, United States, [6]Laboratory for Innovation Science at Harvard, Boston, MA, United States, [7]Department of Physics, University of Toronto, Toronto, ON, Canada, [8]Center for Astrophysics, Harvard & Smithsonian, Harvard University, Cambridge, MA, United States, [9]College of Education, University of Otago, Dunedin, New Zealand

*CORRESPONDENCE
Jessica Sarah Schonhut-Stasik
✉ jessica.s.stasik@vanderbilt.edu

†These authors have contributed equally to this work

**Introduction:** Historically, astronomy has prioritized visuals to present information, with scientists and communicators overlooking the critical need to communicate astrophysics with blind or low-vision audiences and provide novel channels for sighted audiences to process scientific information.

**Methods:** This study sonified NASA data of three astronomical objects presented as aural visualizations, then surveyed blind or low-vision and sighted individuals to elicit feedback on the experience of these pieces as it relates to enjoyment, education, and trust of the scientific data.

**Results:** Data analyses from 3,184 sighted or blind or low-vision survey participants yielded significant self-reported learning gains and positive experiential responses.

**Discussion:** Results showed that astrophysical data engaging multiple senses could establish additional avenues of trust, increase access, and promote awareness of accessibility in sighted and blind or low-vision communities.

## 1 Introduction

### 1.1 Astronomy in the visual

Light is the dominant data source in the Universe; therefore, our sense of sight pervades historical astronomy. For millennia, humans explored the sky with the unaided eye until the invention of the telescope (1608) provided a deeper view of the cosmos. In 1851, the first daguerreotype of a Solar eclipse captured the Sun's light, the first astronomical image (Figure 1). In the twentieth century, the quality and quantity of astronomical data experienced massive growth; during World War II, the development of "false color" enhanced astronomical image interpretability (Mapasyst, 2019), and digital data capture with charge-coupled devices (CCDs; Rector et al., 2015) created the potential to collect tremendous amounts of data.

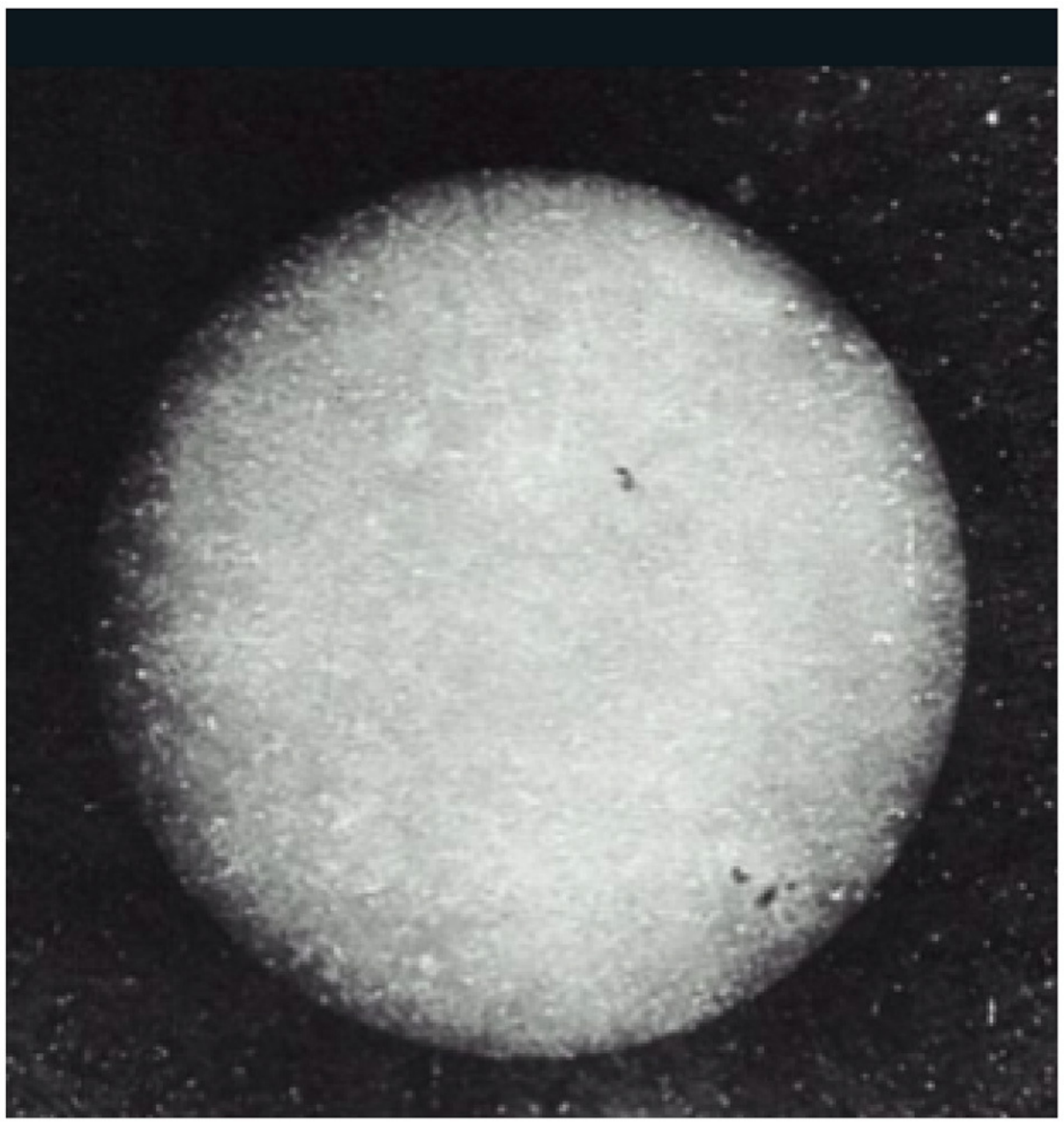

FIGURE 1
First image of the Sun, taken in Fizeau and Foucault (1845). Image from ©ESA at: https://www.esa.int/ESA_Multimedia/Images/2004/03/First_photo_of_the_Sun_1845.

Newly launched space-based observatories captured many of these large data sets, which revolutionizing astronomy's cultural impact (Taterewicz, 1998) by recording the energy, time, and location of photons emitted from cosmic objects. Space telescope data, delivered through NASA's Deep Space Network,[1] is vast and beyond typical human comprehension, from all-sky stellar mapping to imaging gargantuan galaxies (Marr, 2015; Arcand and Watzke, 2017). Images are central to understanding the scope and significance of such vast catalogs (Smith et al., 2011; Arcand et al., 2019), creating an information landscape where archival data have immense value and visualized data hold additional interpretive utility (Hurt et al., 2019).

In the late twentieth century, CCDs began to specialize, observing and transposing light outside visible wavelengths, extending the observable Universe once more. Here, data visualization became essential. NASA's "Great Observatories" [originally the Hubble Space Telescope, Compton Gamma Ray Observatory, Chandra X-ray Observatory, and Spitzer Space Telescope (National Aeronautics and Space Administration, 2009), and now the James Webb Space Telescope] rely on the conversion of their digital transmissions into images (Rector et al., 2015). These data can be combined into aesthetically beautiful images and shared with non-experts (English, 2016), relaying important scientific messages. With the internet, astronomers may couple these innovatively visualized data with worldwide public dissemination (Rector et al., 2015).

1 https://www.nasa.gov/directorates/heo/scan/services/networks/deep_space_network/about

Image aesthetics is essential in aiding data comprehension (Arcand et al., 2013; Rector et al., 2015), and the nature of the visualization varies depending on the audience; for example, a plot communicating a scientific result in a journal or a scientifically accurate but aesthetically pleasing image to communicate publicly (DePasquale et al., 2015). Astronomers must alter images for light outside the visual spectrum regardless of the audience; therefore, it is feasible to use other methods of data vivification (Sturdevant et al., 2022) to elicit new understanding or meaning-making.

Studies on astronomical image processing (Rector et al., 2007, 2017; DePasquale et al., 2015), most notably the use of color (Smith et al., 2011, 2015, 2017; Arcand et al., 2013), have shown image creators must explain the translation process for non-expert audiences and that lay explanations perpetuate the public's confidence in the image's scientific nature and authenticity, even when they depict components concealed from human eyes, indicating the same is possible when translating visual data to sound.

## 1.2 Sonification overview

Sonification is often defined as mapping data to sound to represent information using non-speech audio, the sonic counterpart to visualization (Kramer et al., 2010; Sawe et al., 2020). For decades, image sonification has communicated spatial information to blind or low-vision (BLV) individuals (Meijer, 1992; Yeo and Berger, 2005; Zhang et al., 2014). It is also used for audio alerts (i.e., Geiger counters; Kramer et al., 2010), research (i.e., studying brain wave changes; Parvizi et al., 2018), communications and education (i.e., multimedia cosmology sonification; Ballora and Smoot, 2013), increasing accessibility (i.e., the sonification of visual graphs; Ali et al., 2020), and for art and entertainment (i.e., compositional works based upon neurological or cosmic particle data; Sinclair, 2012).

In recent years, sonification has become more present in the world of astronomy as communicators and researchers alike try to understand how best to engage the BLV community and allow astronomers to conduct their research using sound alone; see Harrison et al. (2022) and Noel-Storr and Willebrands (2022) for discussions on this topic. Generally, a primary goal of astronomy sonification relates to public engagement (Zanella et al., 2022); however, research investigating active listening by astronomers for data exploration and analysis exists (Alexander et al., 2010; Diaz-Merced, 2013). A notable project in this area is the Audible Universe Project by Misdariis et al. (2022), which aims to create a dialogue between sonification and astronomy. Sonified celestial objects include pulsars, the cosmic microwave background, and solar eclipses (McGee et al., 2011; Ballora, 2014; Eclipse Soundscapes, 2022). In the past decade, astronomy-related sonifications have increased dramatically (Zanella et al., 2022), with more than 98 projects globally representing a third of the Sonification Archive (Lenzi et al., 2021).

The most common approaches to sonification are audification and parameter mapping. With audification, data are translated and mapped directly to audio, so the relevant frequencies fall within hearing range; for example, the sonification of gravitational wave signals recorded by the Laser Interferometer Gravitational-Wave Observatory (LIGO) collaboration (Abbot, 2016); which interprets space-time fluctuations caused by passing gravitational waves as an audio signal, allowing us to hear black hole mergers in real-time (Gravitational Wave Open Science Center, 2022).

For parameter mapping, aspects of the data control specific audio parameters. For example, a star's brightness fluctuations or hues of color may control the frequency so that unique features can be identified (Astronify, 2022); it is common to map brightness to pitch as our ears are more sensitive to variations in frequency than volume. The inverse spectrogram is the most common image mapping technique (Sanz et al., 2014), mapping one axis to time and the other to frequency, allocating a corresponding brightness for each pixel to control volume. For example, we may scan an image from left to right, with different pitches indicating the vertical position of objects. Due to the wide variety of possible mappings, an explanation of the process is vital for the listener to extract meaningful information.

Sonification provides astronomy communicators a new avenue to engage the public, particularly BLV communities, traditionally excluded from engagement. Furthermore, sonification is advantageous as a research tool (with or without visually presented information) because listening to data exploits the auditory system's exquisite sensitivity to pattern variation over time, whether perceived as discrete rhythms or changing pitch (Walker and Nees, 2011). In addition, because sounds are multidimensional, we may encode many parallel data streams by mapping each to a different audio dimension (pitch, volume, timbre) or control multiple simultaneous audio streams so our ears can either listen holistically or focus on one stream at a time (Fitch and Kramer, 1994). Finally, we can render each layer of a multi-wavelength image as a separate audio stream (a different note or instrument) to explore the relationship between wavelengths of data. As such, sonification has excellent potential for stimulating curiosity, increasing engagement, and creating an emotional connection with data.

In recent years, NASA has released several sonification projects, showcasing several decades of data (National Aeronautics and Space Administration, 2020, 2022; and others). In 2020, NASA's Chandra X-ray Observatory launched 'The Universe of Sound,[2] providing bespoke audio representations of astronomical datasets for non-expert audiences and working with BLV representatives to create and test sonifications. In this work, we analyze survey data to investigate the effectiveness of sonifications, particularly for BLV communities. This paper represents the first study to explore responses to astronomical sonifications from the BLV community and compare these responses to the experiences of sighted participants.

### 1.3 Researcher perspective/positionality

Before we discuss the results of our work, it is essential to acknowledge our positionality within the context of this study. For some authors, our motivation to explore this topic is shaped by a personal connection to the disability community through lived experience (either in the BLV community or the broader disability community); for others, the motivation lies in the desire to explore alternate data vivification processes and understand how to communicate science to the public effectively. Our own experiences have led us to believe sonification is a positive tool for education and research, and we remain mindful of this bias throughout our analysis. Finally, although our team represents a range of perspectives within the disability and astronomy community, we acknowledge that we remain limited by our lived experiences as a group of majority white individuals living in North America and Canada.

2 https://chandra.si.edu/sound/

## 2 Methods

The primary research questions for this study were:

1. How are data sonifications perceived by the general population and members of the BLV community?
2. How do data sonifications affect participant learning, enjoyment, and exploration of astronomy?

There were two secondary research questions:

1. Can translating scientific data into sound help enable trust or investment, emotionally or intellectually, in scientific data?
2. Can such sonifications help improve awareness of accessibility needs that others might have?

### 2.1 Participants

The research participants were a convenience sample of respondents (18 years and older) to an online survey. We solicited participants from websites including Chandra[3] and Astronomy Picture of the Day (APOD),[4] digital newsletters, social media sites such as Facebook and Twitter for Chandra[5] and APOD,[6] and the social media and contacts of the principal investigator (PI). Further distribution occurred through additional social media sharing. The survey was active on SurveyMonkey[7] for 4 weeks beginning February 24, 2021. We note that SurveyMonkey surveys are compatible with assistive software typically used by the BLV community, particularly screen readers and screen magnification.

The Smithsonian Institutional Review Board[8] determined that this survey was exempt research under Smithsonian Directive 606.[9] The survey started with a participant consent form in which choosing to continue with the survey equaled consent. We provided no compensation to survey participants or dissemination partners.

3 http://chandra.si.edu
4 http://apod.nasa.gov
5 @chandraxray
6 @apod
7 https://www.surveymonkey.com/
8 https://www.si.edu/osp/policies/human-subject-research
9 https://stri-sites.si.edu/permits/sd606/SD606.pdf

TABLE 1 Each row in the table describes the basic parameters for each sonification including a link to the data product, the runtime (in suitable units), the types of sounds used, the number of individual components in the sonification, the wavelength range sonified, and the communication goal of the sonification when it was created.

| Sonified astronomical object | Length | Sounds used | No. of pieces | Wavelength range sonified | Progression across image | Goal of sonification |
|---|---|---|---|---|---|---|
| Galactic Center (https://chandra.si.edu/sound/gcenter.html) | 1.04 min per piece. Total Time: 4.16 min | Instruments: Glockenspiel, String, Piano. | 4 Three individual, one composite. | X-ray (Chandra), Optical (Hubble), Infrared (Spitzer). | Left to right | Communicating detectable structures in different wavelength regimes and highlighting the high density and activity that is present near the Galactic Center. |
| Cassiopeia A (https://chandra.si.edu/sound/casa.html) | 42 s for the first five. 21 s for the sixth. Total Time: 3.52 min | Instruments: String section (double bass, cello, viola, and two violins) | 6 Five individual, one composite. | X-ray (by elemental abundance). | Radial—from the center outwards on four paths. | Revealing the chemical emissions throughout the debris field and highlighting the remnant's shape and structure. |
| Chandra Deep Field South (https://chandra.si.edu/sound/cdf.html) | Total Time: 48 s | Synthetic sounds | 1 | X-ray (by low, medium, high energies). | Bottom to top | Demonstrating the extensive range of X-ray energies/frequencies and demonstrating black hole number density. |

## 2.2 Sonifications

We chose three sonifications from the six available at the time of the study at NASA's Universe of Sound website, choosing those that best represent the collection available for their variation of instrumental vs. synthetic sounds and how they track the visual data—left to right, top to bottom, or radially. Survey participants were presented with the sonifications and their accompanying videos to experience as they were able, followed by short text descriptions (screen-reader adaptable) for each of the represented astronomical objects (the Galactic Center, Cassiopeia A, and the Chandra Deep Field South). The sonifications played in the same order, without counterbalancing, starting with the sonification that used non-synthesized sounds, followed by two more complex sonifications. We provide the details of each sonification at the companion GitHub[10] and highlight the key points, along with links to the sonifications, in Table 1.

## 2.3 Procedure

Our survey began with five demographic questions[11] (age, gender, education level, self-rated knowledge of astronomy, and whether the participant identified as BLV).

10 https://github.com/Jesstella/a_universe_of_sound

11 Breakdowns of demographic information collected for the survey can be found in the Appendix.

Participants were then asked to experience the sonifications and after each, respond to a set of statements using a Likert scale. Each statement began with:

> *"Please respond to this item using a scale of 1 (Disagree Strongly) to 5 (Agree Strongly):"*

The five statements were:

1. I enjoyed this experience.
2. I learned more about the [title of sonification, i.e., Galactic Center] through this experience.
3. Hearing the sounds enhanced my experience.
4. Watching the videos enhanced my experience (if applicable).
5. I trust that this representation is faithful to the science data.

The scale provided the following options: 1 (Disagree Strongly), 2 (Disagree), 3 (Neutral), 4 (Agree), 5 (Agree Strongly).

Following this section, the participants were asked about their overall experience, first:

> *"List up to three words to describe your emotional response to these data sonifications."*

Then, they were asked to rate the following three statements, using the same Likert scale from the first section:

1. *After listening to these data sonifications, I am motivated to listen to more.*

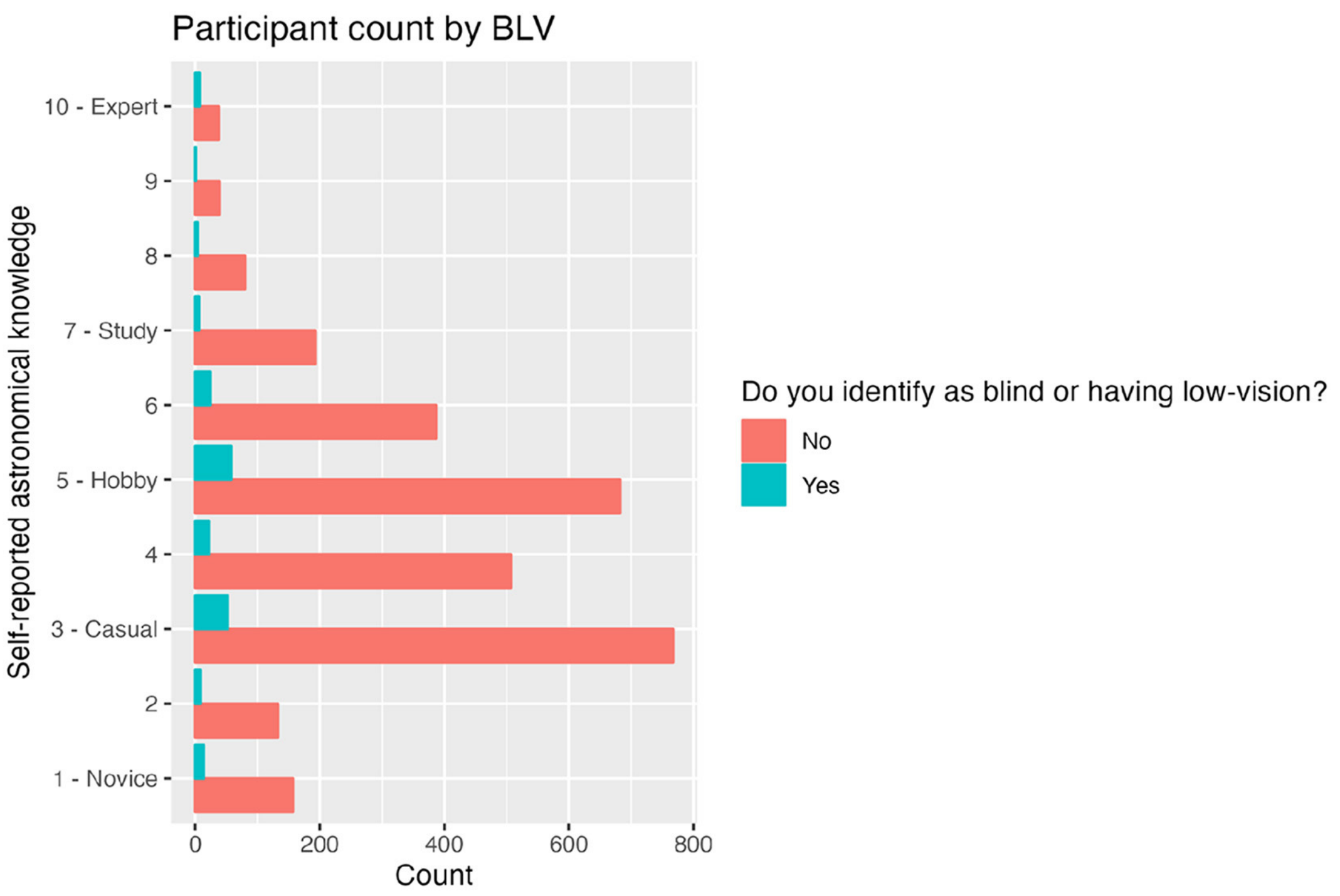

FIGURE 2
Histogram showing the relative amount of BLV and sighted participants. Red (blue) bars represent sighted (BLV) participants. The x-axis shows the participant counts and the y-axis shows the self-reported level of astronomy knowledge from novice to expert. Note the significant difference between BLV and sighted participants. Thirteen participants omitted astronomical knowledge or BLV status and were omitted from this figure.

2. *After listening to these data sonifications, I am interested in learning more about our Universe.*
3. *After listening to these data sonifications, I want to learn more about how others access information about the Universe.*

Finally, they were asked two open-ended questions (which allowed for full sentences). These were:

1. *What recommendations do you have to help the scientific community create better listening experiences?*
2. *If the person who created these data sonifications were here, what question would you ask them?*

Once the survey closed, we exported the data and cleaned and analyzed the 4,346 responses using Python. We removed the entry of one participant who took from March until July 2021 to complete the survey and all responses in which participants did not indicate whether they were BLV or sighted or answered fewer than three non-demographic questions. This cleaning ensured we could compare the results of the BLV and sighted groups for those who engaged with the sonification questions. We removed identical entries by comparing Internet Protocol (IP) addresses and demographic questions. For repeat entries, we kept only the most recent response. Cleaning yielded 3,184 participant responses. See the Appendix for the demographic breakdown of the cleaned sample.

Figure 2 displays self-reported knowledge of astronomy, divided by the BLV (blue) and sighted (red) participants. The apparent contrast in size of the two demographic groups is discussed in Results (Section 3) and Future Work (Section 6.1).

Regarding additional demographics, we note a slight majority of male-identifying participants (57.1%). Participant ages spanned from 18 to 24 years (21.6%) to 65 years and older (16.3%); there is a slight predominance of younger participants, but all age groups are represented at above 10% of the total. Likewise, the self-reported education level of participants ranges from those who completed some of high school to those with advanced postgraduate degrees (i.e., doctorate, LLB, or MD); however, those who completed some of high school were the least represented group (3.5%), with most participants (61.3%) having completed an undergraduate degree or higher. We refer the reader to the tables in the Appendix for a complete breakdown of participants' demographic data.

# 3 Results

## 3.1 Survey question results

Figures 3–6 display responses to the sonification prompts, separated into BLV and sighted participants, and displayed in order of sonification from left to right (Galactic Center, Cassiopeia A, and the Deep Field). We performed 2-sided Kolmogorov–Smirnov (K-S) tests on each set of distributions using the Python module Scipy's `ks_2samp`[12] function. We elected to use the K-S Test, a non-parametric test, because we did not expect the distribution of

12 https://docs.scipy.org/doc/scipy/reference/generated/scipy.stats.ks_2samp.html

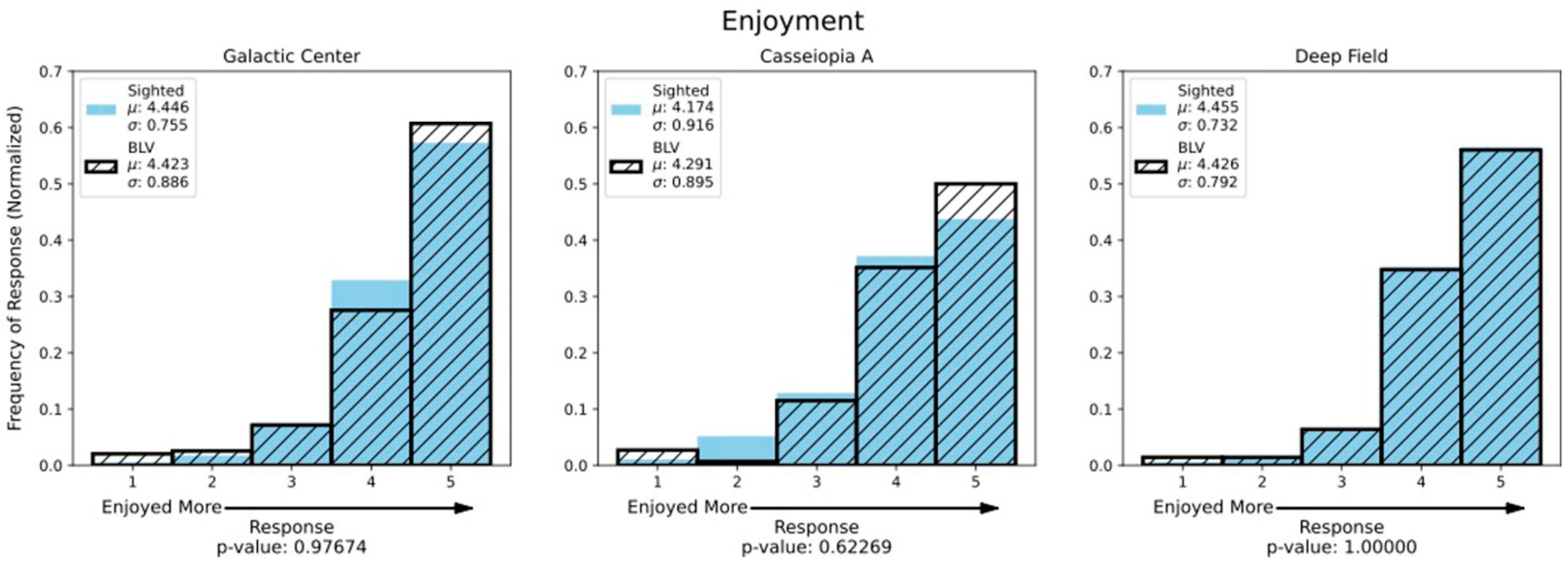

FIGURE 3
These histograms depict the participant's ratings of their enjoyment of each sonification. A rating of one represents the least enjoyment (Disagree Strongly), and five represents the most amount of enjoyment (Agree Strongly). We normalized the histograms for easier comparison between the sighted and BLV groups due to the significant difference in sample size. Blue histograms represent the sighted group, and the black hatched histograms represent the BLV group. **(Left)** Enjoyment rating for the Galactic Center. **(Middle)** Enjoyment rating for Cassiopeia A. **(Right)** Enjoyment rating for the Deep Field.

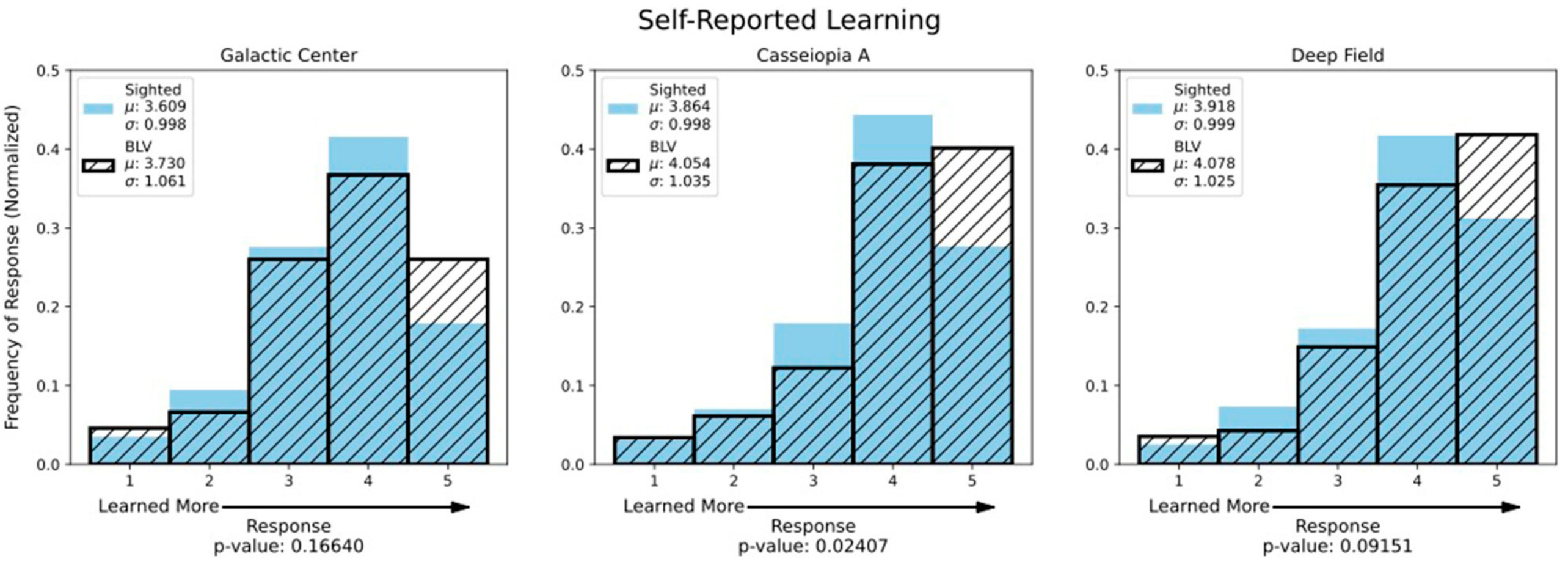

FIGURE 4
These histograms depict the participants' feelings on how much they learned from each sonification. Unless otherwise stated, these and all subsequent histograms follow the same conventions as Figure 3.

responses to our survey to be normal, which visual inspection of the data confirmed. We define a $p$-value of $< 0.05$ as evidence against the null hypothesis and a $p$-value $< 0.001$ as strong evidence against the null hypothesis. Although our sample sizes differ between the BLV and sighted groups, the 2-sided K-S test can accommodate these differences while maintaining validity, and the default "auto" parameter used can handle small sample sizes. However, this sample has more significant uncertainty due to the smaller number of BLV participants. The results of the K-S tests and p-values for all responses can be found in Table 2.

Figure 3 shows participant ratings for the prompt: "I enjoyed this experience." Generally, all participants reported enjoying the sonifications, with the majority selecting 4 (Agree) and 5 (Agree Strongly). A higher number of BLV participants selected 5 for the Galactic Center and Cassiopeia A, whereas the enjoyment ratings for the Deep Field are almost identical for both groups. Cassiopeia A shows the most extensive range of ratings, and although more BLV participants selected the highest rating, the $p$-value does not suggest a statistically significant (0.620) difference between the two groups; however, the $p$-value is significantly lower than for the other sonifications (0.997 and 1.000, respectively).

Ratings for the prompt: "I learned more about the [title of sonification, i.e., Galactic Center] through this experience," are shown in Figure 4. In general, most participants felt they learned something about the cosmic sources. For each sonification, the BLV group rated learning more, particularly for Cassiopeia A, where the low $p$-value ($p = 0.02407$) suggests a statistically significant difference in the responses of the BLV and sighted groups. On average, both groups claimed to learn most about the Deep Field.

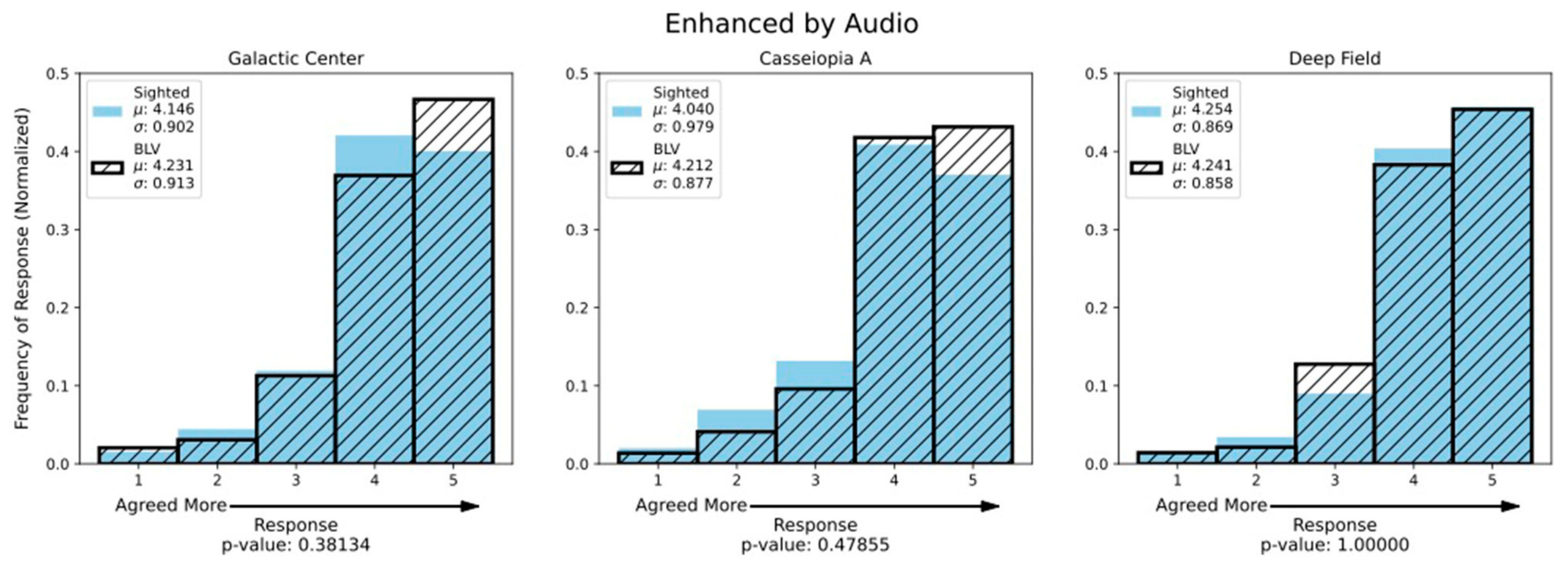

FIGURE 5
These histograms depict the participants' agreement on whether their experience of the images was enhanced by adding sound.

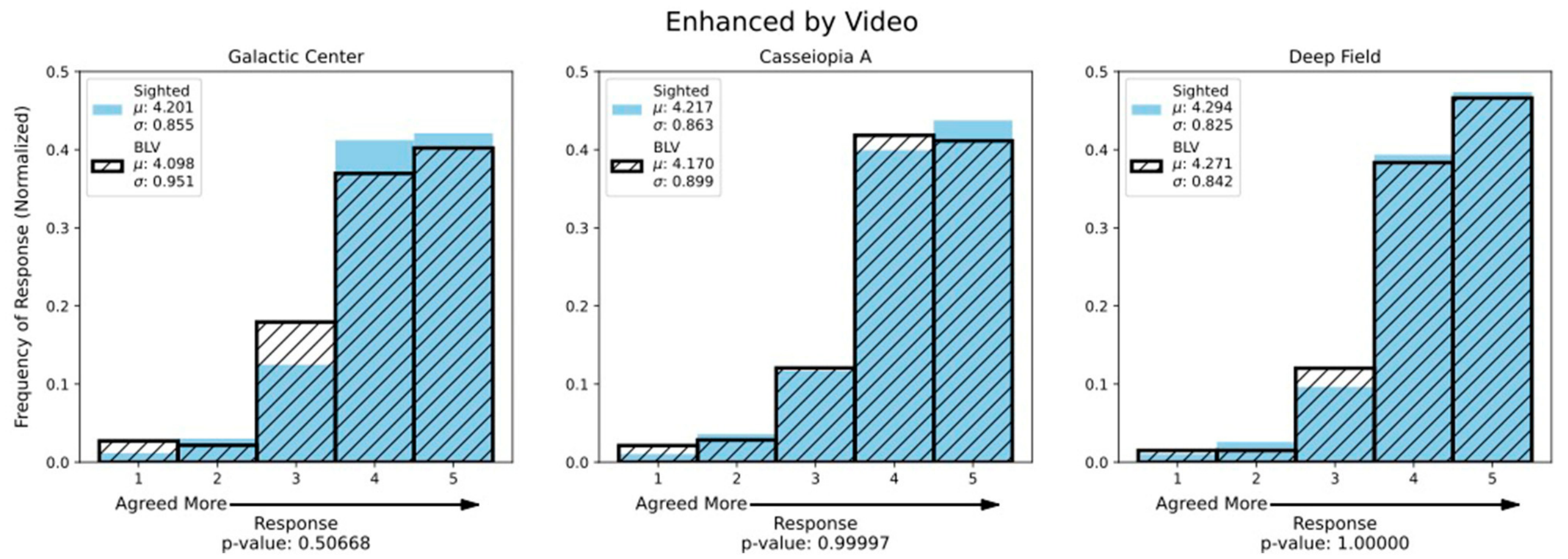

FIGURE 6
These histograms depict the participants' rating of whether their experience was enhanced by watching the included video.

Figure 5 shows ratings for the prompt: "Hearing the sounds enhanced my experience." Generally, participants felt that adding audio to the astronomical images enhanced their experience, particularly for the Deep Field. Interestingly, all the $p$-values are high, suggesting that both the sighted and BLV participants found their experience enhanced to the same extent.

The responses to the prompt "Watching the video enhanced my experience" are depicted in Figure 6. The K-S test results for all three objects indicate that the responses from both groups are statistically similar, with the $p$-values for Cassiopeia A and the Deep Field suggesting the highest similarity. We note that across all three objects, both groups infrequently responded that they disagreed (1 or 2), with 4 and 5 being the most common response, suggesting both groups generally found the video to be a beneficial addition to their experience.

The prompt: "I trust that this representation is faithful to the science data," is displayed in Figure 7. The frequency of 4 and 5 ratings indicating agreement suggests that participants believed the sonifications were scientifically accurate. The high $p$-values (all $> 0.9$) suggest no evidence that the trust levels differed between the groups.

Figure 8 shows the ratings for three prompts given after listening to the sonifications: "I am motivated to listen to more [sonifications]," "I am interested in learning more about our Universe," and "I want to learn more about how others access information about the Universe." The $p$-value for the left-hand histogram implies no difference between the groups regarding whether they wanted to listen to more.

The distribution of ratings regarding interest in continued learning about the Universe and how others access this information differed significantly between the groups ($p$-values $< 0.05$).

TABLE 2 K-S test statistics and *p*-values for the sighted and BLV groups' responses to the survey prompts.

| Prompt | K-S test statistic | *P*-value |
|---|---|---|
| I enjoyed this experience | | |
| Galactic Center | 0.0343 | 0.9767 |
| Cassiopeia A | 0.0625 | 0.6227 |
| Deep Field | 0.0094 | 1.0 |
| I learned more about the (title of sonification, i.e., Galactic Center) through this experience | | |
| Galactic Center | 0.0812 | 0.1664 |
| Cassiopeia A | 0.1249 | **0.0241** |
| Deep Field | 0.1063 | 0.0915 |
| Hearing the sounds enhanced my experience | | |
| Galactic Center | 0.0662 | 0.3183 |
| Cassiopeia A | 0.0704 | 0.4786 |
| Deep Field | 0.0239 | 1.0 |
| Watching the videos enhanced my experience (if applicable) | | |
| Galactic Center | 0.0616 | 0.5067 |
| Cassiopeia A | 0.0261 | 1.0 |
| Deep Field | 0.0177 | 1.0 |
| I trust that this representation is faithful to the science data | | |
| Galactic Center | 0.0390 | 0.9315 |
| Cassiopeia A | 0.0284 | 0.9998 |
| Deep Field | 0.0275 | 0.9999 |
| | | |
| After listening to these data sonifications, I am motivated to listen to more. | 0.0966 | 0.2156 |
| After listening to these data sonifications, I am interested in learning more about our Universe. | 0.1258 | **0.0470** |
| After listening to these data sonifications, I want to learn more about how others access information about the Universe. | 0.1313 | **0.0337** |

Bold *p*-values indicate those at <0.05, representing evidence against the null hypothesis, indicating a difference between the two populations.

In both cases, BLV participants responded 5 (Agree Strongly) more frequently than sighted users. Although sighted participants responded 5 with a lower frequency to these two prompts, the most common responses were still in agreement (4 and 5), indicating that sighted participants were also interested in learning more. Of the 2,203 sighted respondents to these questions, 1,708 responded in agreement to the prompt regarding being motivated to listen more, 1,798 responded in agreement to the prompt about being interested in learning more about the Universe, and 1,710 responded in agreement to the prompt regarding wanting to learn more about how others access information about the Universe. One sighted participant who answered the other two final prompts did not respond to the prompt about wanting to learn more about the Universe.

## 3.2 Word cloud

Figure 9 shows a word cloud,[13] displaying the terms participants used to respond to, "List up to three words to describe your emotional response to these data sonifications." Word size corresponds to their frequency of use. For terms that pertain to a positive experience, the number of instances is as follows: The combined terms "curious" and "curiosity" totaled 329 instances, and "calm" showed the highest number of entries for a single term (326), followed by "interesting" (243). Additional terms included "relaxed/relaxing" (216), "amazed" (169), "wonder" (168), "beautiful" (133), "peaceful" (132), and "awe" (124). Negative terms also appeared, including (but not limited to): "Boring/bored," "confused/confusion," "stress," "disturbed," "pointless," "gimmicky," and "scary," but these appeared with far less regularity.

## 3.3 Open-ended questions

The first open-ended question asked, "What recommendations do you have to help the scientific community create better listening experiences?" There were no character limits imposed on the answers. Using manual inductive coding (Chandra and Shang, 2019) through sampling and re-coding, we collated responses into seven broad categories: general comments, technical, scientific, musical, educational, sensory, and accessibility-related, chosen based on the themes seen in the responses. There were 1,417 responses (removing all non-descriptive responses such as blanks or symbols [i.e., ///]). A complete summary of the responses is available in the data repository.

Amongst the responses, we noted two frequent themes; the first, a common misunderstanding of sonification, both at its conceptual level (e.g., "Include actual sound from space") and in the context of interpretation (e.g., "I don't fully understand the relationship between the sounds and what we are seeing"). These comments suggest an unfamiliarity with sonification as a form of data representation, and the audience may require more background to interpret this representation correctly. We teach students to read graphs and charts visually, so education in sonification might likewise be necessary, echoing the suggestions of Fleming (2023).

The second theme noted is the frequent suggestions regarding the assignment of pitches and other audio parameters to the data, ranging from musical suggestions (e.g., "Please don't stick to the equal temperament system in sound reproduction, so much scientific information is lost or misrepresented that way. Also, why link different things to different pitches suggesting differences in quality better represented by different timbre?") to responses tagged as scientific (e.g., "If you're going to assign a sonification to individual elements, I think you're going to have to find a better way to differentiate between them than to just change the note on the

13 https://www.wordclouds.com/

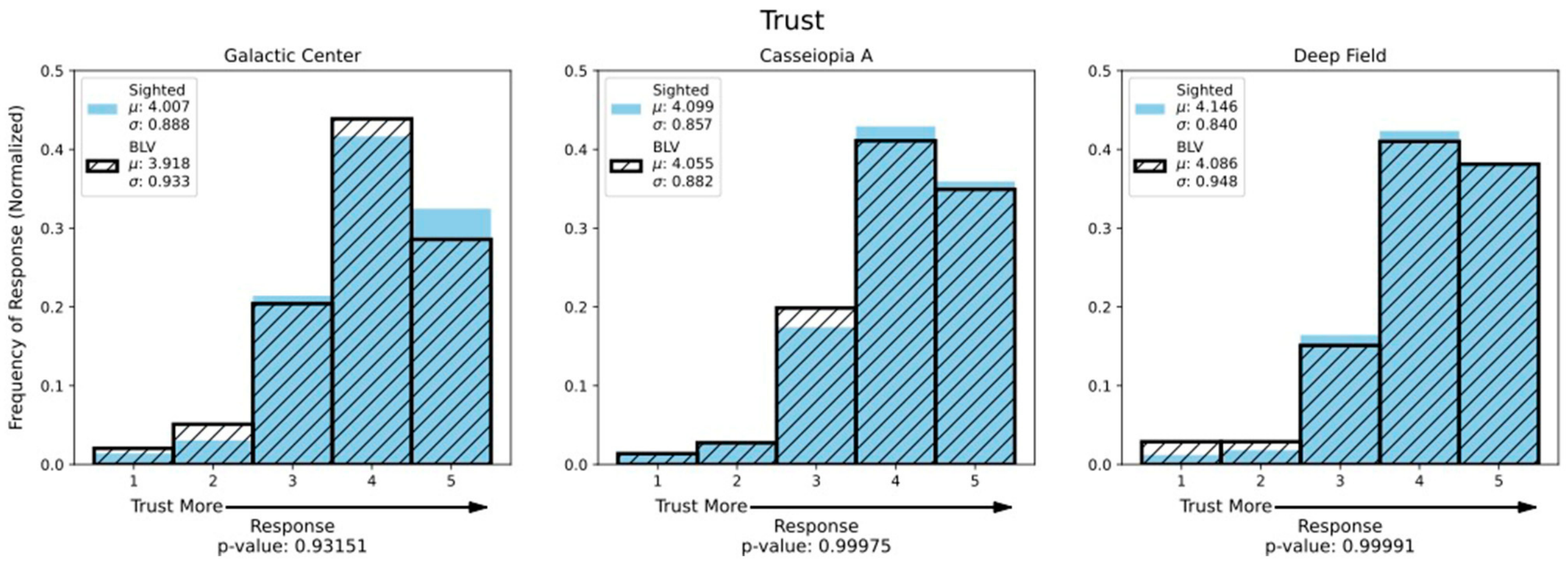

FIGURE 7
These histograms depict the participants' agreement on whether they trusted the sonifications were scientifically accurate.

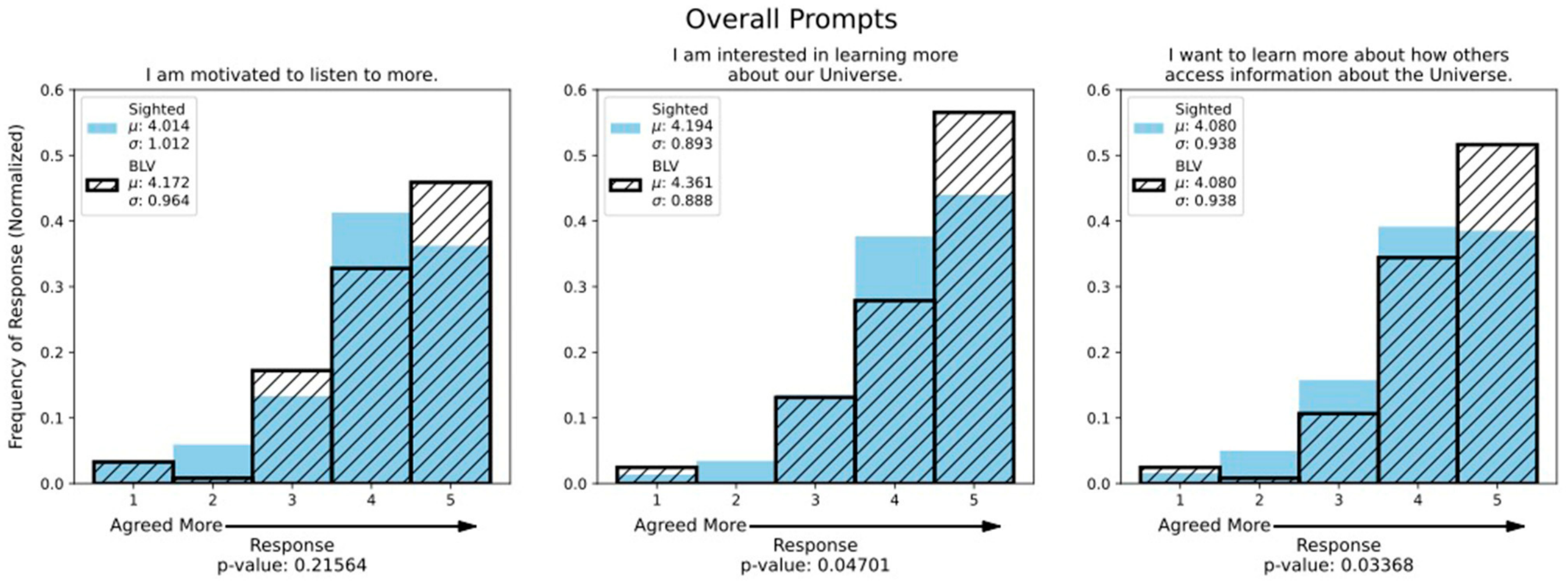

FIGURE 8
These histograms depict ratings for three prompts provided after participants had listened to all sonifications. **(Left)** Ratings for the prompt "After listening to these data sonifications, I am motivated to listen to more." **(Middle)** Ratings for the prompt "After listening to these data sonifications, I am interested in learning more about our Universe." **(Right)** Ratings for the prompt "After listening to these data sonifications, I want to learn more about how others access information about the Universe." Apart from the prompts addressed (a different question for each histogram as opposed to a different sonification), the histogram conventions follow the same as those previous.

scale. Maybe brainstorm a way to differentiate between them based on atomic weight or outer electron shell, assigning a sonification to the sounds the orbits might make."). These responses reflect the question of standardization in sonification, much the same as the standards for visual data representation: how we represent different images or data types in a way that is both interpretable across different sonifications and auditorily pleasant. These standards could improve the feasibility of sonification education.

The second open-ended question asked, "If the person who created these data sonifications were here, what question would you ask them?" We coded these responses into the same seven categories. There were 1,656 responses after removing non-descriptive responses. Across the categories, many questions involved the purpose of these sonifications (e.g., "is the goal enchanting soundscapes or information transfer or enhancing information acquisition in non-visual." and "By a glimpse to photo we can all have these information at once. So what is the use of this?"). Other questions inquired how the audio parameters were mapped to the image data (e.g., "Did you select the frequency distributions to try and make the sonifications tuneful, or are they evenly (linearly, logarithmically) spaced across the audio spectrum?"). The ubiquity of these questions indicates the relative novelty of and lack of familiarity with sonification as a data representation tool among general audiences.

# 4 Discussion

**Enjoyability**

Participants across both groups rated their experience as enjoyable (Figure 3), with slightly higher ratings from the BLV group for the Galactic Center and Deep Field. The word cloud also demonstrates enjoyment (Figure 9), as the majority of words skew toward positive responses (i.e., "peaceful," "wonder," and "relaxing"). It is encouraging that many participants enjoyed the sonifications, demonstrating the benefit of accessible data even to sighted individuals. Figure 8 demonstrates participant interest in hearing more and suggests the benefits of sonification beyond a learning tool for the BLV community to a general engagement tool. It also presents an opportunity to engage sighted groups regarding accessibility in astronomy.

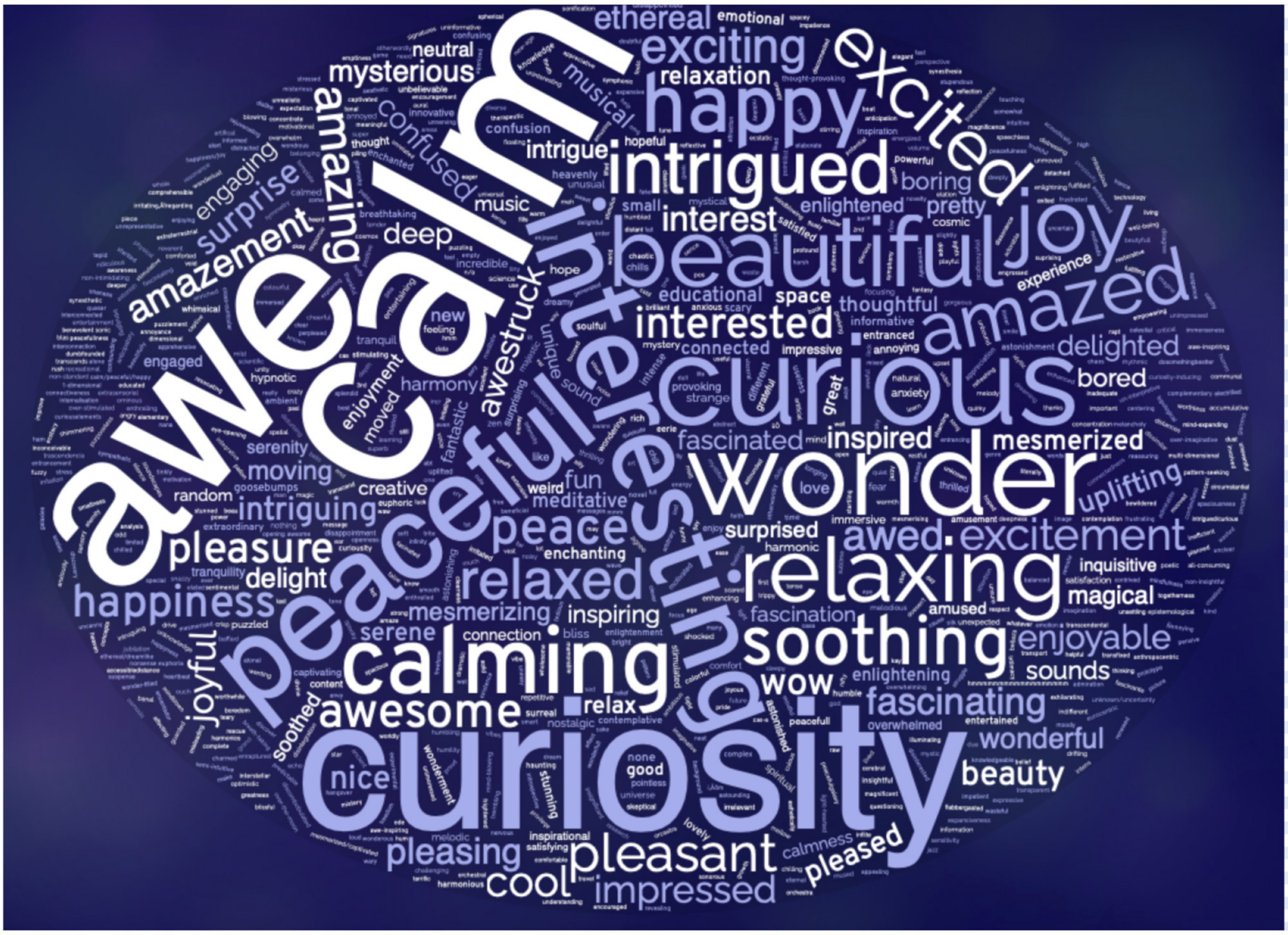

FIGURE 9
This word cloud shows words provided by participants when asked to describe the sonifications. Larger words correspond to those used more frequently.

Surprisingly, the Deep Field showed the widest range of enjoyment, which was unexpected as we reduced the image resolution by a factor of four before being sonified to produce more audible, consistent tones. We made this change to add musical regularity, designed to increase enjoyment. This sonification may have been the least popular as it was the shortest and contains synthetic sounds. Considering this alongside the first open-ended question, which demonstrates participant preference for orchestral sounds, a preference for instrumental sonifications could be demonstrated. In addition, demographic information may be pertinent here; for example, do particular listeners prefer orchestral sounds?

### Learning

The BLV participants reported learning more than the sighted participants (Figure 4), which we expected as some of this group may have lacked exposure to astronomical data due to the nature of this visual science. Generally, this result suggests that sound adds a layer to the experience (or creates an experience) that BLV participants rarely encounter. Furthermore, both groups reported learning more about these objects, suggesting that sound, when added to visual data, can improve self-reported learning regardless of sight, demonstrating the benefit of accessible learning models (see also Figure 3). Interestingly, participants reported learning the least about the Galactic Center, perhaps because it is a more generally known astronomical "object." However, we should note that the Galactic Center was the first sonification heard, which may have affected self-reported learner ratings.

### Enhanced experience

Figure 5 reinforces our finding that accessibility benefits all; the majority of participants found their experience enhanced with the addition of sound. Intuitively, we expected the BLV participants to find more significant enhancement from the audio; however, the similarity in responses between the two groups could reflect the spectrum of sight loss within the BLV community. Among the legally blind population, ∼10–15% have no vision.[14] The remaining 85–90% experience sight loss ranging from light perception to the inability to read text and see images without significant magnification. We did not request BLV participants report their degree of sight loss, so we do not know how many could see the images. Compounding this difficulty is that "visually impaired" among the BLV population has no widely agreed-upon definition, so our BLV sample could include individuals with sight better than the legal cutoff for blindness; however, it seems reasonable to assume that many of our BLV participants could see some of the visualizations included, accounting for the similarities in the groups' responses.

**Trust**

Many participants felt that the sonifications represented scientifically accurate data (Figure 7), and although encouraging, we must be mindful of potential bias. A level of trust may exist due to NASA's association with the project. Interestingly, the level of trust did not vary much between the groups, implying that an accompanying visual component did not increase trust. This result represents the only set of ratings for which the BLV group chose 4 (Agree) to a greater degree than 5 (Agree Strongly) across all sonifications, possibly indicating a critical area of future improvement; if enjoyment, self-reported learning and enhancement from multiple sensory components are high, perhaps trust is the essential aspect to improve. The BLV community chose a rating of 4 more for the Galactic Center image, a sonification with orchestral mappings. The relatively lower trust ratings for the sonifications from BLV participants might reflect the historical exclusion of this community from astronomy and the sciences more broadly.

**Accessibility**

The BLV participants wanted to listen to further sonifications, learn more about the process, and learn more about how others access information about the Universe (Figure 8). They stated agreement for these prompts more consistently than the sighted group. This difference in ratings indicates that our BLV group found their exposure to sonification rewarding, allowing them to learn more about the Universe through a novel method with which they may not have experience. This increased interest from BLV participants could represent a personal investment, supporting their community's requirement for accessible educational materials. Sighted participants rated these prompts with less enthusiasm. Still, they showed a positive trend toward interest, a promising sign that they felt motivated to learn more about information accessibility, potentially increasing awareness for the disabled community. The majority of participants indicated an ongoing interest in sonification following exposure to our study, aligning with our findings of enjoyment (Figure 3), self-reported learning (Figure 4), and feelings that sound enhanced the experience (Figure 5).

With the exception of the self-reported learning from the Cassiopeia A sonification (Figure 4), these overall prompts regarding motivation to learn more about the Universe and about information access regarding the Universe mark the only results wherein the BVI and sighted responses differ to a statistically significant extent. This signifies that while both visually impaired and sighted participants largely enjoy, trust, report learning from, etc. individual sonifications to a similar degree, sonifications on a larger scale appear to be more motivating to BVI participants than to sighted participants.

**Misconceptions**

Responses to our first open-ended question regarding possible improvement (Section 6.1) revealed two potential misunderstandings. The first misconception is the source of the sound (i.e., the sounds are only representations of the data), which is rectifiable with better explanations of the sonification process. Similar misconceptions may also affect visually represented data, for example, the translation of X-ray data to visually accessible images, where the viewer might conclude that these celestial objects are visible to the human eye (Varano and Zanella, 2023). The second misconception is how sonification represents scientific data. This misconception requires more thought than the first. Misunderstanding how we represent the data echoes feelings of mistrust, perhaps due to sonification's novel approach and the lack of exposure to this technique, remedied through more exposure. It suggests that descriptions of the goals and the creation process should be central and involve careful and considerate communication.

## 4.1 Limitations

This study represents a valuable contribution to accessibility in astronomy; however, it is not as rigorous as desired. We selected participants via a convenience sample, where they voluntarily chose to complete our survey after receiving the link from a newsletter (Chandra or APOD) or astronomy-related social media account. Due to the voluntary nature of participation, those involved may be more interested in astronomy and have a base of knowledge, possibly affecting their interpretation of the sonifications. By formulating a questionnaire that (in part) attempts to obtain opinions on sonification products produced by the authors, we may have introduced a social desirability bias, potentially causing participants to respond more favorably to the sonifications. A complete analysis of this effect is outside this work's scope, but we may consider it more thoroughly in future publications.

Our most significant limitation was the lack of BLV respondents, with the smaller sample size resulting in increased uncertainty in the distribution of their responses. Finally, our survey includes a United States-heavy participant distribution due to how we circulated the survey.

# 5 Conclusion

Scientists, data processors, and science communicators are failing to reach and communicate with BLV audiences. We should expand our priorities for processing and presenting information beyond images and present new, novel methods for those with and without sight loss to engage with science. The public

14 https://dsb.wa.gov/dispelling-myths

availability of astronomy data does not necessarily equate to the true accessibility and equity of that data, much as providing a sidewalk in a high-traffic area improves pedestrian safety but remains inherently inaccessible and inequitable without thoughtful design (by cutting the curb). This paper offers suggestions on potential means for universal design for learning (Bernacchio and Mullen, 2007) in astronomical data processing to improve access to scientific research.

Translating data into sonifications is similar to translating language; by considering cultural nuance, we can create sounds that retain astronomical information and impart an accessible mode for scientific communication. A key conclusion is that the sighted participants enjoyed, learned, and had their experience of astronomy enhanced by the sonifications to similar levels as the BLV participants. The responses from the BLV community reinforce the need for access, and the responses from the sighted community show the benefit to all. These results are typical when implementing accessible designs. For example, consider moving airport walkways, a requirement of the Americans with Disabilities Act[15] often enjoyed by those without disabilities. Astronomy, at its core, is a visual science and provides a vital example of the necessity of sonified data for educational and outreach purposes; however, the lack of accessible materials for the BLV community is not specific to astronomy. A review of all potential avenues in which sonification could play an important role is outside the scope of this paper; suffice it to say that, at the very least, in all places where primary data representation is visual, there is a place for a sonified counterpart.

Furthermore, when considering our secondary research question, "Can translating scientific data into sound help enable trust or investment, emotionally or intellectually, in scientific data?" we urgently need accessible data to improve trust. Figure 7 (compared to Figures 3–5) and the first open-ended question demonstrate this. As referenced in the discussion (Section 4), both groups show some degree of mistrust that the sonifications accurately represent the scientific data. In some cases, there is a disconnect as to what the content is showing. We can only cultivate trust through consistent, considerate, and accurate communication. The BLV group generally trusted the data less than the sighted group. Without more detailed information on levels of sightedness, it is hard to determine whether this is due to the inability to compare the visual and audio elements or, perhaps, historical evidence for and societal expectation of astronomy as a purely visual endeavor.

The secondary research question, "Can such sonifications help improve awareness of accessibility needs that others might have?" was explored in Figure 7. The responses reflect that exposure to accessible science data enhances knowledge and accessibility to both groups. These results represent the accessibility needs of the BLV community and the willingness and engagement of the sighted community.

As we progress from this work, the long-term potential learning gains for respondents who engage with sonified data is an important consideration. A single exposure to our sonifications and related questions cannot quantify the long-term learning outcomes of the participants; however, this is an important consideration when implementing sonified materials into more formal educational settings, and it is essential to examine whether using multiple methods would reinforce learning outcomes and retention for students.

One more minor but no less critical conclusion is that participants prefer instrumental sonifications over synthesized sounds. This result is significant because the enjoyment and enrichment of the listener is predicated on the listenership, dictated by how many people listen or include sonifications in their communication efforts. Accessibility to astronomy and scientific data, generally, is still in its infancy. Astronomers need an accelerated effort with adequate resources to reach underserved populations. This project is an important step, but many more are needed.

## 5.1 Future work

Future work must focus on the active engagement of BLV participants while recognizing and accommodating the wide range of visual impairments within this non-homogenous group. Efforts could employ different sampling techniques to recruit a larger sample, particularly for a range of BLV individuals with a scope of astronomy familiarity. BLV participants without astronomy familiarity provide insight into how intuitive sonifications are, whereas participants with more familiarity can share how well sonifications match or enhance their understanding of the objects.

We acknowledge that the BLV category spans a broad range of sight loss that this study does not explore or quantify. Future research should ask participants to comment on the usefulness of the images accompanying the sonification as a proxy for measuring their functional vision. Researchers could also collect data on the accessibility software used while completing the survey (e.g., screen magnification, screen readers, Braille displays, and other methods) to understand whether BLV participants access the survey visually or often visually access their computers. Furthermore, one could ask for feedback regarding the visualizations to improve the accessibility of these data representations to those with low vision.

Astronomy communicators must continue to address and resolve misunderstandings of the sonification process by improving accompanying descriptions of the techniques used. These updates must consider the lens of trust in science and be mindful of creating minimal opportunities for miscommunication. To understand this better, we must capture data on the number of times a participant plays a sonification, providing a more objective measure of comprehensibility, intuitiveness, enjoyment, and a desire to understand.

Further studies could gauge the self-reported knowledge of music and technology. Many participants gave feedback on the musical quality, indicating an understanding of music theory, and many also gave technical feedback (bearing in mind that some technical proficiency is required to access the survey).

Although we collected participants' ages, we did so primarily to compare the representativeness of our sample to the overall U.S. population (see Table 3 in the Appendix) and provide

15 Americans With Disabilities Act of 1990, Pub. L. No. 101-336, 104 Stat. 328 (1990).

thorough information regarding our participants. Future work could explore whether age correlates with enjoyment, self-reported learning, trust, and overall responses to the sonifications, although analysis of this is beyond the scope of our work. We could also explore the role of misconceptions with age. Future studies should be mindful that some participants may have hearing loss, which we do not report here, and could impact the response to sonifications. Hearing loss is more likely with increased age and could further impact the relationship between age and response to sonifications. Other demographic questions, in particular self-reported knowledge of astronomy, could also reveal interesting relationships with responses to sonification and can be explored in the future.

Finally, this work could extend to investigate actual learning outcomes, as opposed to self-reported learning (as in this study). However, this is outside this paper's scope and would involve a participant and control group learning with and without access to sonification.

Input from the broader community is invaluable, and we are encouraged by the recommendations received and excited to implement them into new work. We look forward to collaborating with others throughout astronomy and related fields to make as much data available to as many people as possible. Additional resources are available for this paper on a companion GitHub (see text footnote 10) and a frozen Zenodo[16] repository.

## Data availability statement

The original contributions presented in the study are included in the article/Supplementary material, further inquiries can be directed to the corresponding author.

## Ethics statement

The studies involving humans were approved by Smithsonian Institutional Review Board. The studies were conducted in accordance with the local legislation and institutional requirements. The participants provided their written informed consent to participate in this study, by way of a consent form at the beginning of the survey.

## Author contributions

KA: Writing—original draft, Writing—review & editing. JS-S: Writing—original draft, Writing—review & editing. SK: Writing—original draft, Writing—review & editing. GS: Writing—original draft, Writing—review & editing. MR: Writing—original draft. MW: Writing—original draft. BH: Writing—original draft. LS: Writing—review & editing.

16 https://zenodo.org/record/8248153

## Funding

The author(s) declare financial support was received for the research, authorship, and/or publication of this article. This paper was written with funding from NASA under contract NAS8-03060 with the principal investigator working for the Chandra X-ray Observatory. NASA's Marshall Space Flight Center manages the Chandra program. The Smithsonian Astrophysical Observatory's Chandra X-ray Center controls science and flight operations from Cambridge and Burlington, Massachusetts. Additional support for the sonifications came from NASA's Universe of Learning (UoL). UoL materials are based upon work supported by NASA under award number NNX16AC65A to the Space Telescope Science Institute, with Caltech/IPAC, Jet Propulsion Laboratory, and the Smithsonian Astrophysical Observatory.

## Acknowledgments

The authors gratefully acknowledge their colleagues at the Center for Astrophysics, NASA and particularly NASA's Astronomy Picture of the Day, for their gracious dissemination help with the study. JS-S acknowledges the Frist Center for Autism and Innovation in the School of Engineering at Vanderbilt University, who fund the Neurodiversity Inspired Science and Engineering Graduate Fellowship. SK acknowledges the Marshall Scholarship.

## Conflict of interest

The authors declare that the research was conducted in the absence of any commercial or financial relationships that could be construed as a potential conflict of interest.

## Supplementary material

The Supplementary Material for this article can be found online at: https://www.frontiersin.org/articles/10.3389/fcomm.2024.1288896/full#supplementary-material